\documentclass[reprint,noshowpacs,noshowkeys,prd,balancelastpage,nofootinbib]{revtex4}
\usepackage{amsfonts}
\usepackage{mdframed}
\usepackage{amssymb}
\usepackage{footnote}
\usepackage{amsmath}
\usepackage{graphicx}
\usepackage{float}
\usepackage[font={footnotesize,it}]{caption}
\usepackage[utf8]{inputenc}
\usepackage{natbib}
\usepackage{tikz}
\usepackage{tikz-3dplot}
\usepackage[colorlinks=true,
            linkcolor=purple,
            urlcolor=purple,
            citecolor=blue]{hyperref}

\begin{document}

\title{Minimum asymptotic energy of a charged spherically symmetric
thin-shell wormhole and its stability}
\author{S. Habib Mazharimousavi}
\email{habib.mazhari@emu.edu.tr}
\affiliation{Department of Physics, Faculty of Arts and Sciences, Eastern Mediterranean
University, Famagusta, North Cyprus via Mersin 10, Turkiye}

\begin{abstract}
The static spherically symmetric thin shell of proper mass $m$ and electric
charge $q$ around a central star or black hole of mass $M$ is in general
unstable. Although with fine-tuned $m,q,$ and $M$ it can be considered
quasi-stable. These facts have been recently proved by Hod in \cite{HOD1}
from which we are inspired to investigate the stability status of the throat
of a spherically symmetric thin-shell wormhole (TSW) with the same proper
mass and charge. Precisely, in this compact paper, we introduce the radius of
stability for the throat of a static spherically symmetric charged TSW, by
optimizing its energy measured by an asymptotic observer. Furthermore, we
prove that at the stable equilibrium radius, the tension on the throat is
zero, and a perturbation in the form of an initial kinetic energy results in
an effective attractive potential such that the throat oscillates around the
equilibrium point.
\end{abstract}

\date{\today }
\pacs{}
\keywords{Thin-shell wormhole; Charged; Stable; }
\maketitle

\section{Introduction}

In \cite{MANN} the static spheres - consist of test neutral massive
particles - formed around a static spherically symmetric black hole
spacetime have been studied. According to \cite{MANN} the equation of
geodesics for a massive test particle in the vicinity of a static
spherically symmetric central object such as a black hole or a star with the
line element 
\begin{equation}
ds^{2}=-f\left( r\right) dt^{2}+\frac{dr^{2}}{g\left( r\right) }+r^{2}\left(
d\theta ^{2}+\sin ^{2}\theta d\phi ^{2}\right) ,  \label{N1}
\end{equation}%
is expressed by%
\begin{equation}
\dot{r}^{2}+g\left( r\right) \left( 1+\frac{l^{2}}{r^{2}}-\frac{E^{2}}{%
f\left( r\right) }\right) =0.  \label{N2}
\end{equation}%
Herein, $l=g_{\phi \phi }\dot{\phi}$ and $E=-g_{tt}\dot{t}$ are the orbital
angular momentum and the energy per unit mass of the particle which are both
conserved. Furthermore, the angular velocity of the particle with respect to
an asymptotic observer at rest is found to be%
\begin{equation}
\Omega =\frac{d\phi }{dt}=\frac{lf\left( r\right) }{Er^{2}}.  \label{N3}
\end{equation}%
Considering geodesics with zero angular velocity implies $l=0,$ since $%
f\left( r\right) >0$ outside a black hole or star. Therefore, (\ref{N2}) for
a particle with zero angular velocity becomes%
\begin{equation}
\dot{r}^{2}+g\left( r\right) \left( 1-\frac{E^{2}}{f\left( r\right) }\right)
=0.  \label{N4}
\end{equation}%
Now, we ask the main question: Can a test massive particle with zero angular
velocity remain at rest at a finite distance from the event horizon of a
black hole or from the surface of a star? To answer this question, we refer
to (\ref{N4}) where the behavior of the effective potential i.e., 
\begin{equation}
V_{eff}\left( r\right) =g\left( r\right) \left( 1-\frac{E^{2}}{f\left(
r\right) }\right) ,  \label{N5}
\end{equation}%
is crucial. Considering an equilibrium radius i.e., $r=R,$ where $\dot{r}=0$
and $\ddot{r}=0,$ one finds $E=\sqrt{f\left( R\right) }$ and $f^{\prime
}\left( R\right) =0.$ Therefore, for instance in the Schwarzschild black
hole spacetime where $f\left( r\right) =g\left( r\right) =1-\frac{2M}{r},$
the second condition can not be satisfied i.e., $f^{\prime }\left( R\right)
\neq 0$. Similarly, for the Reissner-Nordstr\"{o}m (RN) black hole spacetime
with $f\left( r\right) =g\left( r\right) =1-\frac{2M}{r}+\frac{Q^{2}}{r^{2}}$
the equilibrium radius is obtained to be $R=\frac{Q^{2}}{M}$ which is
smaller than the radius of the event horizon $r_{+}=M+\sqrt{M^{2}-Q^{2}}.$
Therefore, there is no such equilibrium radius outside the event horizon for
the RN black hole. However, if the spacetime describes outside a static star
of mass $M,$ and electric charge $Q$ such that $Q>M$ and the radius of the
massive star is smaller than $R=\frac{Q^{2}}{M}$, then such an equilibrium
radius exists. Moreover, at the equilibrium radius, one finds $%
V_{eff}^{\prime \prime }\left( R\right) >0$ which implies the equilibrium in
this configuration is stable. Moreover, in \cite{MANN}, it was proved that,
unlike the RN black hole, the black holes in Einstein-Nonlinear electrodynamics theory permit stable equilibrium radius for the massive test
particles.

Following the seminal work of  Wei, Zhang, Liu, and Mann \cite{MANN}, in 
\cite{HOD1} Hod answered another, yet relevant, question in the same
context: "\textit{Going beyond the test particle approximation, is it
possible to place a static thin Dyson shell \cite{Dyson} around central
compact objects in the standard Einstein-Maxwell field theory}?" The answer,
however, was NO in general but for the fine-tuned cases YES. In \cite{HOD1}
the radius of the thin shell surrounding the central object is set to $R$ so
that the inner and outer line elements describing the spacetime are given,
respectively, by 
\begin{equation}
ds^{2}=-\left( 1-\frac{2M}{r}\right) dt^{2}+\frac{dr^{2}}{1-\frac{2M}{r}}%
+r^{2}\left( d\theta ^{2}+\sin ^{2}\theta d\phi ^{2}\right) ,\text{ for }%
r\leq R,  \label{I1}
\end{equation}%
and%
\begin{equation}
ds^{2}=-\left( 1-\frac{2E\left( R\right) }{r}+\frac{q^{2}}{r^{2}}\right)
dt^{2}+\frac{dr^{2}}{1-\frac{2E\left( R\right) }{r}+\frac{q^{2}}{r^{2}}}%
+r^{2}\left( d\theta ^{2}+\sin ^{2}\theta d\phi ^{2}\right) ,\text{ for }%
R\leq r.  \label{I2}
\end{equation}%
Herein, $M$ is the asymptotic mass of the central object, $E\left( R\right) $
is the asymptotic energy of the thin shell and $q$ is its electric charge.
The line element of the thin shell itself is also given by%
\begin{equation}
ds_{shell}^{2}=-d\tau ^{2}+R^{2}\left( d\theta ^{2}+\sin ^{2}\theta d\phi
^{2}\right) ,  \label{I3}
\end{equation}%
in which $\tau $ is the proper time and $r=R$ is the radius of the shell.
Applying the Israel junction conditions \cite{I1,I2} at $r=R$, one finds the
proper mass of the thin shell i.e., $m=4\pi R^{2}\sigma $ with $\sigma $
being the surface energy density given by 
\begin{equation}
\sigma =\frac{1}{4\pi R}\left( \sqrt{1-\frac{2M}{R}}-\sqrt{1-\frac{2E\left(
R\right) }{R}+\frac{q^{2}}{R^{2}}}\right) .  \label{I4}
\end{equation}%
Consequently, the proper energy of the shell in terms of $M,m,q$, and $R$ is
obtained to be (see also \cite{HOD2})%
\begin{equation}
E\left( R\right) =-\frac{m^{2}}{2R}+\frac{q^{2}}{2R}+m\sqrt{1-\frac{2M}{R}}.
\label{I5}
\end{equation}%
As was introduced in \cite{HOD1}, the static equilibrium radius of the thin
shell is the critical point(s) of the asymptotic energy (\ref{I5}). Hence by
solving $\frac{dE\left( R\right) }{dR}=0$ one obtains 
\begin{equation}
R_{eq}=\frac{2M\left( q^{2}-m^{2}\right) ^{2}}{\left( q^{2}-m^{2}\right)
^{2}-4M^{2}m^{2}}.  \label{I6}
\end{equation}%
Furthermore, the second derivative reveals that 
\begin{equation}
\left. \frac{d^{2}E\left( R\right) }{dR^{2}}\right\vert _{R=R_{eq}}<0,
\label{I7}
\end{equation}%
which indicates $R=R_{eq}$ is an unstable equilibrium radius. This
instability implies that the shell after a small perturbation either
collapses toward the central object or expands toward infinity. The rate of
collapsing or vaporizing of the shell after a radial perturbation depends
strongly on the values of $q,m,$ and $M.$ Upon setting $\frac{q}{\sqrt{%
m\left( m+M\right) }}\rightarrow 1^{+}$, the rate is so small that it takes
infinite proper time for the shell to complete the process of collapsing or
blowing. Hence the shell is practically considered to be quasi-stable.

Let us recall that in general relativity, similar to thin shells that
connect two incomplete spacetimes, thin-shell wormholes (TSW) also connect
two spacetimes with a different mechanism. Originally TSW was introduced by
Matt Visser in \cite{MV1,MV2}. The mechanical stability of TSWs has always
been one of the main concerns. The general structure of the stability
analysis of TSWs was initially established by Poisson and Visser in \cite%
{MV3} where they introduced the mechanical stability of the TSW in the
Schwarzschild black hole bulk spacetime.

The mathematical and physical analogy between the thin shells and TSWs
suggests applying the formalism introduced by Hod in \cite{HOD1} to the
TSWs. Therefore, the main question that we shall find an answer to in this
compact paper is: \textit{Is there a stable equilibrium radius for a static and
spherically symmetric TSW of proper mass }$m$\textit{\ and charge }$q$%
\textit{?} In what follows we prove that the answer to this question is a
definite YES.

\section{The field equations and the solutions}

Let us start with a static spherically symmetric TSW whose static throat is
a timelike thin shell of radius $a$, proper mass $m$ and electric charge $q.$
The line element on the $2+1-$dimensional throat is given by%
\begin{equation}
ds^{2}=-d\tau ^{2}+a^{2}\left( d\theta ^{2}+\sin ^{2}\theta d\phi
^{2}\right) ,  \label{1}
\end{equation}%
in which $\tau $ is the proper time and $a$ is its radius. On the other
hand, the asymptotic observers on either side of the throat express the line
element of the $3+1$-dimensional spacetimes that are connected by the throat
is in the form 
\begin{equation}
ds^{2}=-\left( 1-\frac{2E\left( a\right) }{r}+\frac{q^{2}}{r^{2}}\right)
dt^{2}+\frac{dr^{2}}{1-\frac{2E\left( a\right) }{r}+\frac{q^{2}}{r^{2}}}%
+r^{2}\left( d\theta ^{2}+\sin ^{2}\theta d\phi ^{2}\right) ,  \label{2}
\end{equation}%
in which $E\left( a\right) $ is the asymptotic energy and $q$ is the
electric charge of the throat. We note that the asymptotic energy $E\left(
a\right) $ and the proper mass of the throat $m$ are not equal although they
are related. In the sequel, we shall find this relation.

According to the Israel junction formalism, the spacetime across the throat
has to be continuous everywhere, particularly at the location of the throat
which implies%
\begin{equation}
\left( \frac{dt}{d\tau }\right) ^{2}=\frac{1}{1-\frac{2E\left( a\right) }{a}+%
\frac{q^{2}}{a^{2}}}.  \label{3}
\end{equation}%
Furthermore, the second fundamental form is discontinuous such that 
\begin{equation}
\left[ k_{i}^{j}\right] -\left[ k\right] \delta _{i}^{j}=8\pi S_{i}^{j},
\label{4}
\end{equation}%
in which $k_{i}^{j},$ and $k=k_{i}^{i}$ are the mixed extrinsic curvature
and its trace, respectively, and $S_{i}^{j}=diag\left( -\sigma ,p,p\right) $
is the surface energy-momentum tensor describing the fluid of proper mass $%
m=4\pi a^{2}\sigma $ and charged $q$ on the throat. Applying (\ref{4}) one
finds (for detail see \cite{E1,E2})%
\begin{equation}
\sigma =-\frac{1}{2\pi a}\sqrt{1-\frac{2E\left( a\right) }{a}+\frac{q^{2}}{%
a^{2}}},  \label{5}
\end{equation}%
and 
\begin{equation}
p=\frac{1}{4\pi a}\frac{1-\frac{E\left( a\right) }{a}}{\sqrt{1-\frac{%
2E\left( a\right) }{a}+\frac{q^{2}}{a^{2}}}}.  \label{6}
\end{equation}%
Next, by equating the proper mass $m$ and $4\pi a^{2}\sigma ,$ we calculate
the asymptotic energy of the TSW given by%
\begin{equation}
E\left( a\right) =-\frac{m^{2}}{8a}+\frac{q^{2}}{2a}+\frac{a}{2}.  \label{7}
\end{equation}%
The expression of the asymptotic energy of the TSW contains three terms. The
term $-\frac{m^{2}}{8a}$ stands for the gravitational self-energy, $\frac{%
q^{2}}{2a}$ is the electromagnetic self-energy and $\frac{a}{2}$ is the
structural energy which is non-zero even with $m=q=0.$ Next, we minimize the
asymptotic energy in terms of the radius of the throat. In other words we
solve $\frac{dE\left( a\right) }{da}=0$ to obtain 
\begin{equation}
a=a_{eq}=\frac{1}{2}\sqrt{4q^{2}-m^{2}}.  \label{8}
\end{equation}%
To have the equilibrium radius physical one should impose $4q^{2}>m^{2}$ and
accordingly one gets%
\begin{equation}
\left. \frac{d^{2}E\left( a\right) }{da^{2}}\right\vert _{a=a_{eq}}=\frac{1}{%
a_{eq}}>0,  \label{9}
\end{equation}%
which implies that the equilibrium radius is a minimum point for the asymptotic energy, yielding a stable equilibrium radius. By substituting the
stable radius i.e., 
\begin{equation}
a_{st}=a_{eq},  \label{10}
\end{equation}%
into the surface density (\ref{5}) and the angular pressure (\ref{6}) one
finds 
\begin{equation}
\sigma =\frac{m}{\pi a_{st}}<0,  \label{11}
\end{equation}%
and interestingly $p=0.$ This, in turn, implies that on the throat there is
no tension. Finally, the expression of the asymptotic energy at the stable
radius is obtained to be%
\begin{equation}
E_{st}=E_{\min }\left( a\right) =a_{st},  \label{12}
\end{equation}%
revealing that at such a stable radius the gravitational self-energy and the
electromagnetic self-energy cancel each other resulting in zero tension on
the throat and consequently its stability. This therefore strongly suggests
that uncharged spherically symmetric TSW cannot be stable.

\section{Dynamic stability analysis}

\begin{figure}[tbph]
\includegraphics[width=80mm,scale=1]{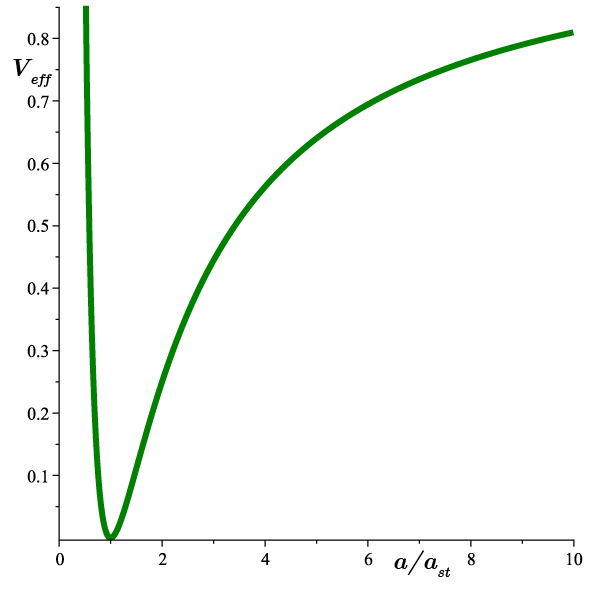}
\caption{The effective potential (\protect\ref{16}) versus $\frac{a}{a_{st}}.
$ This potential implies that the throat, initially located at the
equilibrium radius with $\frac{a}{a_{st}}=1$, will oscillate about this
point upon a perturbation. }
\label{f1}
\end{figure}
In this section, we study the dynamic motion of the TSW whose equilibrium
radius is the stable radius given in Eq. (\ref{8}). To do so we re-express
the induced line element on the throat to be in its dynamic form where $a$
in (\ref{1}) depends on the proper time $\tau $. Accordingly the surface
energy density and the angular pressure on the throat read as \cite{E1,E2}%
\begin{equation}
\sigma =-\frac{1}{2\pi a}\sqrt{1-\frac{2a_{st}}{a}+\frac{q^{2}}{a^{2}}+\dot{a%
}^{2}},  \label{13}
\end{equation}%
and%
\begin{equation}
p=\frac{1}{8\pi a}\frac{2a\ddot{a}+2\dot{a}^{2}+1-\frac{a_{st}}{a}}{\sqrt{1-%
\frac{2a_{st}}{a}+\frac{q^{2}}{a^{2}}+\dot{a}^{2}}},  \label{14}
\end{equation}%
respectively. Eq. (\ref{13}) yields%
\begin{equation}
\dot{a}^{2}+V_{eff}\left( a\right) =0,  \label{15}
\end{equation}%
where 
\begin{equation}
V_{eff}\left( a\right) =\left( 1-\frac{a_{st}}{a}\right) ^{2}.  \label{16}
\end{equation}%
Eq. (\ref{15}) is a one-dimensional equation of motion with the effective
potential $V_{eff}\left( a\right) $. Let us add that, to find (\ref{16}) we
have assumed that the proper mass $m$ of the throat is constant such that $%
\sigma =\frac{m}{4\pi a^{2}}.$ Clearly at $a=a_{st}$ both $\dot{a}$ and $%
V_{eff}\left( a\right) $ vanish, however, a small perturbation in the form
of an initial kinetic energy turns it into%
\begin{equation}
\dot{a}^{2}+V_{eff}\left( a\right) =\dot{a}_{0}^{2},  \label{17}
\end{equation}%
where $\dot{a}_{0}^{2}$ is the initial kinetic energy. In the vicinity of
the equilibrium point the linearized form of Eq. (\ref{17}) is obtained to be%
\begin{equation}
\ddot{a}+\frac{1}{a_{st}^{2}}\left( a-a_{st}\right) ^{2}\simeq 0,  \label{18}
\end{equation}%
where the higher orders have been neglected. The letter implies that the
throat oscillates around its equilibrium radius i.e.,\ $a=a_{st}$ with the
frequency $\omega =\frac{1}{a_{st}}$, that is in agreement with the stable
equilibrium point. Although (\ref{18}) is obtained for small perturbation, the structure of the effective potential (\ref{16}) suggests that for a
strong perturbation the motion is oscillatory as well. In Fig. \ref{f1} we
plotted the effective potential versus $\frac{a}{a_{st}}$ which depicts the
strong stability of the throat with the equilibrium radius $a=a_{st}$.

\section{Conclusion}

Concerning the facts reported by Hod in \cite{HOD1}, \textit{a charged thin
shell of proper mass }$m$\textit{\ and electric charge }$q$\textit{\ formed
around a static spherically symmetric cosmological object such as a black
hole or star of mass }$M$\textit{\ is unstable} \textit{at its equilibrium
radius unless }$\frac{q}{\sqrt{m\left( m+M\right) }}\rightarrow 1^{+}$, we
investigated the stability status of a static spherically symmetric TSW of a
proper mass $m$ and an electric charge $q$. We have proved that, unlike the
thin shell reported in \cite{HOD1}, TSWs are definite stable at their
equilibrium radius. In terms of the proper mass and the electric charge
located at the throat of the TSW, the equilibrium radius has been found to
be $a_{eq}=a_{st}=\frac{1}{2}\sqrt{4q^{2}-m^{2}}$ such that the spacetime
across the throat is described by 
\begin{equation}
ds^{2}=-\left( 1-\frac{2a_{st}}{r}+\frac{q^{2}}{r^{2}}\right) dt^{2}+\frac{%
dr^{2}}{1-\frac{2a_{st}}{r}+\frac{q^{2}}{r^{2}}}+r^{2}\left( d\theta
^{2}+\sin ^{2}\theta d\phi ^{2}\right) .  \label{19}
\end{equation}%
Furthermore, the tension\ or the angular pressure on the throat at the
stable equilibrium radius is zero. We also investigated the dynamics of the
throat after a perturbation in the form of an initial kinetic energy. As
displayed in Fig. \ref{f1}, the effective potential forming around the
equilibrium radius is attractive such that the throat oscillates about the
equilibrium radius. For a weak perturbation, its frequency has been
calculated to be $\omega =\frac{1}{a_{st}}.$

\end{document}